\documentclass[twocolumn,showpacs,preprintnumbers,amsmath,amssymb]{revtex4}

\usepackage{graphicx}
\usepackage{dcolumn}
\usepackage{bm}

\begin{document}

\title{Continuous dynamical protection of two-qubit entanglement from uncorrelated dephasing, bit flipping, and dissipation}

\author{F. F. Fanchini}
 \email{felipe@ifsc.usp.br}

\author{R. d. J. Napolitano}
\affiliation{Instituto de F\'{\i}sica de S\~{a}o Carlos, Universidade de S\~{a}o Paulo, Caixa Postal 369, 13560-970, S\~{a}o Carlos, SP, Brazil}

\date{\today}

\begin{abstract}
We show that a simple arrangement of external fields, consisting of a static component and an orthogonal rotating component, can continuously decouple a two-qubit entangled state from uncorrelated dephasing, bit flipping, and dissipation at finite temperature. We consider a situation where an entangled state shared between two non-interacting qubits is initially prepared and left evolve under the environmental perturbations and the protection of external fields. To illustrate the protection of the entanglement, we solve numerically a master equation in the Born approximation, considering independent boson fields at the same temperature coupled to the different error agents of each qubit.
\end{abstract}


\maketitle

\section{Introduction}

Entanglement of qubits is a fundamental resource of quantum information processing \cite{nielsen00}. A pure entangled state, initially prepared as a superposition of factorized states, inevitably decoheres under environmental noise \cite{zurek91} and could even disentangle completely within a finite time interval \cite{diosi03}. Consequently, any potentially successful design of a device intended to perform quantum computation must anticipate ways to preserve qubit entanglement.

There are very sophisticated methods developed to protect quantum states, like error-correcting codes \cite{shor95} and strategies based on decoherence-free subspaces and subsystems \cite{zanardi97}, where logical qubits do not necessarily coincide with physical qubits. This usually requires that the quantum information corresponding to $N$ logical qubits be stored in more than $N$ physical qubits. It is also possible to protect the quantum information stored directly in physical qubits by using the method of dynamical decoupling \cite{viola98}. We expect that a fully functional quantum computer will use all these protecting tools in a coordinated way for maximum efficiency and fidelity. Here, due to its relative simplicity, we focus on the continuous version of dynamical decoupling \cite{romero04,fanchini07} and show that it can protect a two-qubit entangled state at finite temperature from uncorrelated dephasing, bit flipping, and dissipation.

The paper is organized as follows. Section II presents the total Hamiltonian describing the qubit system and the rest of the universe. We also introduce the general method of dynamical decoupling employed to derive the simple control fields. The master equation for the evolution of the two-qubit density matrix is derived in Sec. III. The description of the environmental errors is given in Sec. IV, where we demonstrate the efficacy of the continuous dynamical decoupling in protecting entanglement. Finally, we conclude in Sec. V.

\section{The total Hamiltonian}

The Hamiltonian for the two-qubit system and the environment is written as
\begin{eqnarray}
H(t)=H_{0}(t)+H_{E}+H_{\rm int},\label{H}
\end{eqnarray}
where $H_{0}(t)$ is the qubit-system Hamiltonian, including terms describing the action of the control fields, $H_{E}$ is the environmental Hamiltonian, and $H_{\rm int}$ is the term representing the interaction between the qubits and their surroundings. The general prescription for dynamical decoupling \cite{viola98,facchi05} consists of finding a control Hamiltonian, acting only on the qubit system, such that its corresponding unitary evolution operator $U_{c}(t)$ is periodic and
\begin{eqnarray}
\int ^{t_{c}} _{0} U^{\dagger}_{c}(t)H_{\rm int}U_{c}(t) dt=0,\label{integral}
\end{eqnarray}
where $t_{c}$ is the period of $U_{c}(t)$. In the following we explain a possible strategy to find a combination of static and simple oscillating external fields leading to a periodic unitary operator $U_{c}(t)$ satisfying Eq. (\ref{integral}).

The interaction Hamiltonian $H_{\rm int}$ describing two qubits coupled to their surroundings is written as
\begin{eqnarray}
H_{\rm int}={\bf B}_{1}\cdot {\bm \sigma}_{1}+{\bf B}_{2}\cdot {\bm \sigma}_{2},\label{Hint}
\end{eqnarray}
where ${\bf B}_{k}=\sum _{m=1} ^{3}B_{k,m}{\bf \hat{x}}_{m}$, for $k=1,2$, with ${\bf \hat{x}}_{1}\equiv{\bf \hat{x}}$, ${\bf \hat{x}}_{2}\equiv{\bf \hat{y}}$, ${\bf \hat{x}}_{3}\equiv{\bf \hat{z}}$, and $B_{k,m}$, for $k=1,2$ and $m=1,2,3$, are operators that act on the environmental Hilbert space. Here, for $k=1,2$, ${\bm \sigma}_{k}={\bf \hat{x}}\sigma_{k,x}+{\bf \hat{y}}\sigma_{k,y}+{\bf \hat{z}}\sigma_{k,z}$, and $\sigma_{k,x}$, $\sigma_{k,y}$, and $\sigma_{k,z}$ are the Pauli matrices acting on qubit $k$. The unitary operator $\exp(-2in_{x}\pi t \sigma_{1,x}/t_{c})$ is periodic with period $t_c$ for any integer $n_{x}\neq 0$. If we identify this operator with $U_{c}(t)$ in Eq. (\ref{integral}) and take $H_{\rm int}$ as given by Eq. (\ref{Hint}), the integration does not give zero in general, but eliminates all terms proportional to $\sigma_{1,y}$ and $\sigma_{1,z}$. If, to replace $U_{c}(t)$ in Eq. (\ref{integral}), we consider the operator $\exp(-2in_{x}\pi t \sigma_{1,x}/t_{c})\exp(-2in_{z}\pi t \sigma_{1,z}/t_{c})$ instead, where $n_{x}$ and $n_{z}$ are non-zero integers, then the integration eliminates all terms proportional to ${\bm \sigma}_{1}$ if $n_{x}\neq n_{z}$, there remaining only the term $t_{c}{\bf B}_{2}\cdot {\bm \sigma}_{2}$. Hence, to satisfy Eq. (\ref{integral}), we choose
\begin{eqnarray}
U_{c}(t)=U_{2}(t)U_{1}(t)=U_{1}(t)U_{2}(t),\label{Uc}
\end{eqnarray}
since ${\bm \sigma}_{1}$ and ${\bm \sigma}_{2}$ commute, where
\begin{eqnarray}
U_{k}(t)=\exp\left(-i \frac{2n_{x}\pi t}{t_{c}}\sigma_{k,x}\right)\exp\left(-i \frac{2n_{z}\pi t}{t_{c}}\sigma_{k,z}\right),\label{Uk}
\end{eqnarray}
for $k=1,2$. Hence, if the control Hamiltonian is written as $H_{c}(t)={\bm \Omega}(t)\cdot\left( {\bm \sigma}_{1}+{\bm \sigma}_{2}\right)$, then Eqs. (\ref{Uc}) and (\ref{Uk}) imply the very simple external-field configuration of our previous work on single-qubit operations, Ref. \cite{fanchini07}:
\begin{eqnarray}
{\bm \Omega}(t)={\bf \hat{x}}n_{x}\omega +n_{z}\omega \left[{\bf \hat{z}} \cos\left( n_{x}\omega t \right)-{\bf \hat{y}} \sin\left( n_{x}\omega t \right) \right] ,\label{Omega}
\end{eqnarray}
where $\omega =2\pi/t_{c}$.

To study the protection of entanglement against environmental sources of noise, we focus on a situation where the two qubits are prepared in an entangled pure state at time $t=0$. We further assume that the qubits do not interact with each other and, if they could be isolated from the rest of the universe, their non-local state would remain unchanged. Thus, let us take $H_{q}=\omega_{0}(\sigma_{1,z}+\sigma_{2,z})$ as the unperturbed two-qubit Hamiltonian, using units for which $\hbar =1$. Let $\tau$ be the time interval during which we intend to preserve the entanglement. We then choose $t_{c}=\tau /N$, where $N$ is an integer, so that $U_{k}(\tau )=I$, for $k=1,2$.

Equation (\ref{H}) gives the Hamiltonian for the evolution of the ket $\left|\Psi (t)\right\rangle $ representing the joint state of the qubits and their environments. For the sake of simplicity, we assume there is a global, static control field along the $z$ axis chosen to cancel $H_{q}$ exactly. Hence, the remaining terms of $H_{0}(t)$ represent the action of additional control fields. We identify the evolution dictated by $H_{0}(t)$ with the action of the unitary operator $U_{c}(t)$ of Eqs. (\ref{Uc}) and (\ref{Uk}).

Here we show that the very simple field configuration of Eq. (\ref{Omega}) can also prevent a two-qubit entangled state from disentangling due to uncorrelated dephasing, bit flipping, and dissipation at finite temperature. We emphasize that Eq. (\ref{Omega}) is a simple combination of a static field along the $x$ axis and a rotating field in the $yz$ plane. Moreover, addressing each qubit independently is not necessary; the field is supposed to be spatially uniform in the neighborhood surrounding both qubits.

\section{The master equation}

In the interaction picture, the Hamiltonian is given by
\begin{eqnarray}
H_{I}(t)=\sum _{k=1} ^{2}\sum _{m=1} ^{3}U^{\dagger}_{E}(t)B_{k,m}U_{E}(t) U^{\dagger}_{c} (t)\sigma_{k,m}U_{c} (t),\label{HI}
\end{eqnarray}
where $\sigma_{k,1}\equiv \sigma_{k,x}$, $\sigma_{k,2}\equiv \sigma_{k,y}$, $\sigma_{k,3}\equiv \sigma_{k,z}$, $U_{E}(t)=\exp(-iH_{E}t)$, and we have used Eq. (\ref{Hint}). The quantities $U^{\dagger}_{c} (t)\sigma_{k,m}U_{c} (t)$, for $k=1,2$ and $m=1,2,3$, are rotations of $\sigma_{k,m}$, whose matrix elements, $R_{m,n}(t)$, are real functions of time:
\begin{eqnarray}
U^{\dagger}_{c} (t)\sigma_{k,m}U_{c} (t)= \sum_{n =1}^{3} R_{m,n}(t)\sigma_{k,n}. \label{rot}
\end{eqnarray}
If we define the operators $E_{k,m}(t)=U^{\dagger}_{E}(t)B_{k,m}U_{E}(t)$, for $k=1,2$ and $m=1,2,3$, and use Eqs. (\ref{HI}) and (\ref{rot}), then the interaction Hamiltonian becomes
\begin{eqnarray}
H_{I}(t)=\sum _{k=1} ^{2}\sum _{m=1} ^{3}\sum_{n =1}^{3}R_{m,n}(t)E_{k,m}(t) \sigma_{k,n}.\label{HI2}
\end{eqnarray}

In the interaction picture, the Redfield master equation describing the temporal evolution of the two-qubit reduced density matrix, $\rho _{I}(t)$, is written as \cite{breuer02}:
\begin{eqnarray}
\frac{d\rho _{I}(t)}{dt}=-\int^{t}_{0}dt^{\prime} {\rm Tr}_{E}\left\{{\left[H_{I}(t),\left[H_{I}(t^{\prime}),\rho _{E}\rho _{I}(t)\right]\right]}\right\},\label{master}
\end{eqnarray}
where we have assumed the noise is low enough that the Born approximation is valid. We also notice that Eq. (\ref{master}) is not a Markovian master equation, since the dynamical-decoupling process occurs in a time scale shorter than the environmental correlation time; this is the reason we keep $t$ as the upper limit of the integral on the right-hand side of Eq. (\ref{master}) (cf. p. 132 of Ref. \cite{breuer02}). Here, $\rho _{E}$ is the initial environmental density matrix, $\rho _{E}=\exp(-\beta H_{E})/Z$, where $Z$ is the partition function given by $Z={\rm Tr}_{E}\left[\exp(-\beta H_{E})\right]$, $\beta =1/k_{B}T$, $k_{B}$ is the Boltzmann constant, and $T$ is the absolute temperature, assumed to be the same in the surroundings of both qubits. By substituting Eq. (\ref{HI2}) into Eq. (\ref{master}) we encounter the quantities ${\rm Tr}_{E}\left[ E_{k,m}(t)\rho_{E}E_{k^{\prime},m^{\prime}}(t^{\prime})\right] $, for $k,k^{\prime}=1,2$ and $m,m^{\prime}=1,2,3$. To illustrate our methodology in a simple manner, we suppose that the reservoir operators at the position of one qubit are uncorrelated with the reservoir operators at the position of the other. Moreover, we also assume the qubits and their respective environments are identical. Thus, we can write ${\rm Tr}_{E}\left[ E_{k,m}(t)\rho_{E}E_{k^{\prime},m^{\prime}}(t^{\prime})\right] =\delta_{k,k^{\prime}}C_{m,m^{\prime}}(t,t^{\prime})$, for $k,k^{\prime}=1,2$ and $m,m^{\prime}=1,2,3$, where $C_{m,m^{\prime}}(t,t^{\prime})$ is the correlation function between components $m$ and $m^{\prime}$ of environmental field operators calculated at the same qubit position. We, thus, define the quantities
\begin{eqnarray}
D_{p,q}(t)=\sum_{m,n=1}^{3}R_{m,p}(t)\int _{0}^{t}dt^{\prime}R_{n,q}(t^{\prime})C_{m,n}(t,t^{\prime}),\label{Dab}
\end{eqnarray}
for $p,q=1,2,3$. Hence, the master equation now becomes
\begin{eqnarray}
\frac{d\rho_{I}(t)}{dt}=\sum_{k=1}^{2}\sum_{p,q=1}^{3}D_{p,q}(t)\left[ \sigma _{k,p},\rho _{I}(t)\sigma _{k,q}\right] \nonumber \\
+\sum_{k=1}^{2}\sum_{p,q=1}^{3}D^{\star}_{p,q}(t)\left[ \sigma _{k,q}\rho _{I}(t),\sigma _{k,p}\right],\label{master2}
\end{eqnarray}
where we have assumed the environmental fields are Hermitian.

\section{Continuous dynamical decoupling}

To solve Eq. (\ref{master2}) we need $D_{p,q}(t)$, Eq. (\ref{Dab}), and, therefore, we must calculate the correlation functions $C_{m,n}(t,t^{\prime})$, for $m,n=1,2,3$. Here, as stated above, we consider independent dephasing, bit flipping, and dissipation. Associated with these errors, we introduce six independent boson fields, three at each qubit position, all of them at the same finite temperature $T$. Accordingly, the terms appearing in Eq. (\ref{Hint}) can be written as
\begin{eqnarray}
{\bf B}_{k}\cdot {\bm \sigma}_{k}=\sigma _{k,z}\sum _{\lambda}\left[ g _{1,\lambda}a_{k,\lambda} +g ^{\star}_{1,\lambda}a^{\dagger}_{k,\lambda}\right]\nonumber \\
+\sigma _{k,x}\sum _{\lambda}\left[ g _{2,\lambda}b_{k,\lambda} +g ^{\star}_{2,\lambda}b^{\dagger}_{k,\lambda}\right]\nonumber \\
+\left(\sigma _{k,x}+i\sigma _{k,y}\right) \sum _{\lambda}g _{3,\lambda}c_{k,\lambda} \nonumber \\
+\left(\sigma _{k,x}-i\sigma _{k,y}\right) \sum _{\lambda}g ^{\star}_{3,\lambda}c^{\dagger}_{k,\lambda} ,\label{baths}
\end{eqnarray}
for $k=1,2$, where $g _{1,\lambda}$, $g _{2,\lambda}$, and $g _{3,\lambda}$ are complex coupling constants that do not depend on the qubit-position index $k$, reflecting the fact that the qubits are surrounded by identical environments, $a_{k,\lambda}$, $b_{k,\lambda}$, and $c_{k,\lambda}$ are, respectively, the annihilation operators for mode $\lambda$ of the boson field associated with dephasing, bit flipping, and dissipation, with respective creation operators $a^{\dagger}_{k,\lambda}$, $b^{\dagger}_{k,\lambda}$, and $c^{\dagger}_{k,\lambda}$. The field operators depend on $k$, since, although identical, the qubit environments are uncorrelated. The only non-zero commutators of these operators are: $[a_{k,\lambda},a^{\dagger}_{k,\lambda}]=1$, $[b_{k,\lambda},b^{\dagger}_{k,\lambda}]=1$, and $[c_{k,\lambda},c^{\dagger}_{k,\lambda}]=1$, for $k=1,2$, and all $\lambda$. Therefore, we take the environmental Hamiltonian as given by $H_{E}=\sum_{k=1}^{2}\sum_{\lambda}\left[ \omega _{1,\lambda}a^{\dagger}_{k,\lambda}a_{k,\lambda}+\omega _{2,\lambda}b^{\dagger}_{k,\lambda}b_{k,\lambda} + \omega _{3,\lambda}c^{\dagger}_{k,\lambda}c_{k,\lambda}\right] $, where $\omega _{1,\lambda}$, $\omega _{2,\lambda}$, and $\omega _{3,\lambda}$ are the $\lambda$th-mode frequencies of the fields associated with dephasing, bit flipping, and dissipation, respectively. Using Eq. (\ref{baths}), we can calculate the correlation functions $C_{m,m^{\prime}}(t,t^{\prime})={\rm Tr}_{E}\left[ E_{1,m}(t)\rho_{E}E_{1,m^{\prime}}(t^{\prime})\right] ={\rm Tr}_{E}\left[ E_{2,m}(t)\rho_{E}E_{2,m^{\prime}}(t^{\prime})\right] $, for $m,m^{\prime}=1,2,3$. The only non-zero correlations are:
\begin{eqnarray}
C_{1,1}(t,t^{\prime})&=&{\cal K}_{2}(t-t^{\prime})+2{\rm Re} \left[ {\cal L}_{2}(t-t^{\prime})\right] \nonumber \\
& &+{\cal K}_{3}(t-t^{\prime})+2{\rm Re} \left[ {\cal L}_{3}(t-t^{\prime})\right] ,\label{C11}\\
C_{1,2}(t,t^{\prime})&=&i{\cal K}_{3}(t-t^{\prime})-2{\rm Im} \left[ {\cal L}_{3}(t-t^{\prime})\right] ,\label{C12}\\
C_{2,1}(t,t^{\prime})&=&-i{\cal K}_{3}(t-t^{\prime})+2{\rm Im} \left[ {\cal L}_{3}(t-t^{\prime})\right] ,\label{C21}\\
C_{2,2}(t,t^{\prime})&=&{\cal K}_{3}(t-t^{\prime})+2{\rm Re} \left[ {\cal L}_{3}(t-t^{\prime})\right] ,\label{C22}\\
C_{3,3}(t,t^{\prime})&=&{\cal K}_{1}(t-t^{\prime})+2{\rm Re} \left[ {\cal L}_{1}(t-t^{\prime})\right] ,\label{C33}
\end{eqnarray}
where the complex functions ${\cal K}_{m}(t)$ and ${\cal L}_{m}(t)$, for $m=1,2,3$, are given by
${\cal K}_{m}(t)=\sum_{\lambda}\left|g_{m,\lambda}\right|^{2}\exp(i\omega_{m,\lambda}t)$ and ${\cal L}_{m}(t)=\sum_{\lambda}\left|g_{m,\lambda}\right|^{2}\exp(i\omega_{m,\lambda}t)/\left[ \exp(\beta \omega_{m,\lambda})-1\right] $.

In the limit in which the number of environmental normal modes per unit frequency becomes infinite, we define spectral densities for $m=1,2,3$ as $J_{m}(\omega)=\sum _{\lambda}\left|g_{m,\lambda}\right|^{2}\delta (\omega -\omega _{m,\lambda})$, with $\omega \in [0,+\infty)$ and interpret the summations in ${\cal K}_{m}(t)$ and ${\cal L}_{m}(t)$ as integrals over $\omega$: ${\cal K}_{m}(t)=\int ^{\infty}_{0}d\omega J_{m}(\omega )\exp(i\omega t)$ and ${\cal L}_{m}(t)=\int ^{\infty}_{0}d\omega J_{m}(\omega)\exp(i\omega t)/[\exp(\beta  \omega )-1]$. For our present purpose of illustrating protection against SDE, it suffices to assume ohmic spectral densities with the same cutoff frequency $\omega _{c}$, that is, $J_{m}(\omega)=\eta _{m}\omega \exp(-\omega/\omega _{c})$, where $\eta _{m}$, for $m=1,2,3$, are dimensionless constants giving the respective strengths of dephasing, bit flipping, and dissipation. Calculating the continuum versions of ${\cal K}_{m}(t)$ and ${\cal L}_{m}(t)$, using the ohmic spectral densities, gives ${\cal K}_{m}(t)=\eta _{m} \omega _{c}^{2}/\left(1-i\omega _{c}t\right)^{2}$ and ${\cal L}_{m}(t)=(\eta _{m} /\beta ^{2}) \Psi ^{(1)}\left(1+1/(\beta  \omega _{c})-it/\beta \right) $, where $\Psi ^{(1)}$ is the first polygamma function. By substituting these results into Eqs. (\ref{C11}), (\ref{C12}), (\ref{C21}), (\ref{C22}), and (\ref{C33}), we obtain the correlations that appear in Eq. (\ref{Dab}), where the rotation matrix elements are obtained from Eqs. (\ref{Uc}), (\ref{Uk}), and (\ref{rot}). Once we have the coefficients $D_{p,q}(t)$, for $p,q=1,2,3$, then we can solve Eq. (\ref{master2}) numerically.

\begin{figure}
\includegraphics[width=9cm]{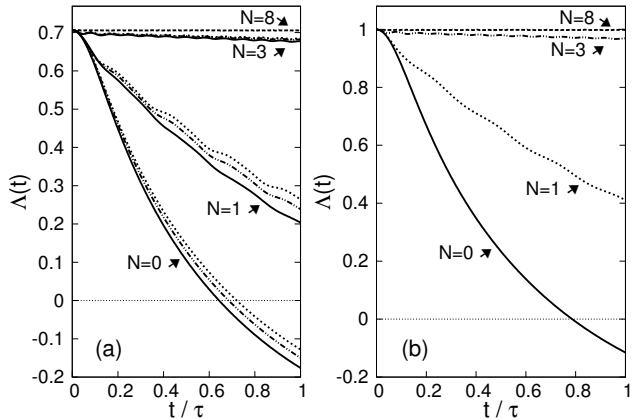}
\caption{\label{figure1} $\Lambda (t)$ for the evolution of different initial entangled states. The concurrence as a function of time is given by $\max \{ 0,\Lambda (t)\} $. (a) $\Lambda (t)$ for initial states with $\Lambda (0)\approx0.70711$, when the control fields are turned off (indicated with $N=0$) and on ($N=1,3,8$). The solid and dotted lines correspond to the initial states of Eq. (\ref{phi}), with $\theta =\pi /8$ and $\theta =3\pi /8$, respectively. The double-dot-dashed line corresponds to the initial states of Eq. (\ref{psi}), with $\theta =\pi /8$ and $\theta =3\pi /8$ giving the same $\Lambda (t)$. (b) $\Lambda (t)$ for initial states with $\Lambda (0)=1$ (Bell states). The initial states of Eq. (\ref{phi}) evolve resulting in the same $\Lambda (t)$ for $\theta =\pm \pi /4$. Also, the initial states of Eq. (\ref{psi}) evolve resulting in the same $\Lambda (t)$ for $\theta =\pm \pi /4$. Here we show the results for Eq. (\ref{phi}), since they differ from those of Eq. (\ref{psi}) only in the fourth decimal place. The solid line represents the result for the control field turned off, while the dotted, double-dot-dashed, and dashed lines show the results for control fields with $N=1,3,8$, respectively.}
\end{figure}
\begin{figure}
\includegraphics[width=9cm]{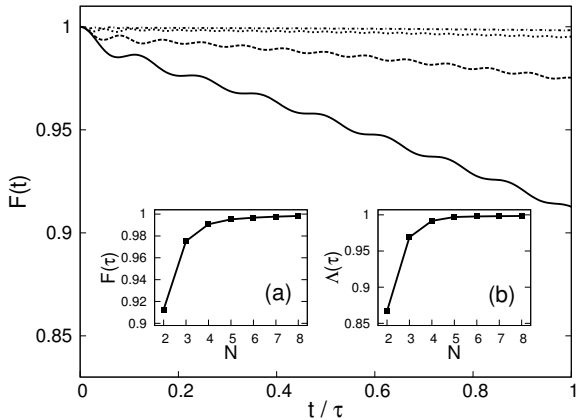}
\caption{\label{figure2}  Fidelity as a function of time for Bell states. The solid, dashed, dotted, and dot-dashed lines give the fidelities when the control fields are turned on, using $N=2,3,5,8$, respectively. The insets (a) and (b) show, respectively, the fidelity and concurrence, both calculated at $t=\tau $, as functions of $N$.}
\end{figure}

Here we impose the extreme situation where the disentanglement occurs faster than the time scale defined by the inverse of the cutoff frequency, $1/\omega _{c}$, and the thermal correlation time, $\tau_{B}=\beta/\pi$. Hence, we take $\tau =2\pi/\omega _{c}$ and, therefore, $t_{c}=\tau /N$, as discussed below Eq. (\ref{Uk}). For concreteness, in the numerical calculations we choose $n_x=2$, $n_z=1$, $\tau =10^{-10}$s, a temperature of $T=0.1$K, $\eta _{1}=1/16$, $\eta _{2}=1/64$, and $\eta _{3}=1/256$. As initial pure states, we consider two classes:
\begin{eqnarray}
\left| \Phi (\theta)\right\rangle =\cos \theta \left|\uparrow\uparrow\right\rangle+\sin \theta \left|\downarrow\downarrow\right\rangle ,\label{phi}\\
\left| \Psi (\theta)\right\rangle =\cos \theta \left|\uparrow\downarrow\right\rangle+\sin \theta \left|\downarrow\uparrow\right\rangle ,\label{psi}
\end{eqnarray}
for $\theta \in [0,2\pi )$, where we adopt the usual spin notation.

To measure entanglement we use the concurrence \cite{wootters98}, defined as the maximum between zero and $\Lambda (t)$, with
\begin{eqnarray}
\Lambda (t) =\lambda _{1}-\lambda _{2}-\lambda _{3}-\lambda _{4},
\end{eqnarray}
where $\lambda _{1} \geq \lambda _{2} \geq \lambda _{3} \geq \lambda _{4}$ are the square roots of the eigenvalues of the matrix $\rho (t)\sigma _{1,y}\sigma _{2,y}\rho ^{\star}(t)\sigma _{1,y}\sigma _{2,y}$, where $\rho ^{\star}(t)$ is the complex conjugation of $\rho (t)$, the reduced density matrix of the two qubits in the Schr\"{o}dinger picture. Thus, if $\Lambda (t)$ gets less than or equal to zero, there is no entanglement and the state is separable. Our aim is to use external fields to keep $\Lambda (t)$ fixed. Figure \ref{figure1} shows $\Lambda (t)$ for the evolution of different initial entangled states. In panel (a), $\Lambda (t)$ is shown for initial states with $\Lambda (0)\approx0.70711$, when the control fields are turned off (indicated with $N=0$) and on ($N=1,3,8$). The solid and dotted lines correspond to the initial states of Eq. (\ref{phi}), with $\theta =\pi /8$ and $\theta =3\pi /8$, respectively. The double-dot-dashed line corresponds to the initial states of Eq. (\ref{psi}), with $\theta =\pi /8$ and $\theta =3\pi /8$ giving the same $\Lambda (t)$. In panel (b) $\Lambda (t)$ is given for Bell initial states, for which $\Lambda (0)=1$. The initial states of Eq. (\ref{phi}) evolve resulting in the same $\Lambda (t)$ for $\theta =\pm \pi /4$. Also, the initial states of Eq. (\ref{psi}) evolve resulting in the same $\Lambda (t)$ for $\theta =\pm \pi /4$. In this figure panel, we show the results for Eq. (\ref{phi}), since they differ from those of Eq. (\ref{psi}) only in the fourth decimal place. The solid line represents the result for the control field turned off, while the dotted, double-dot-dashed, and dashed lines show the results for control fields with $N=1,3,8$, respectively.

From the general theory of dynamical decoupling \cite{viola98,facchi05}, the protection gets better as $N$ gets larger. Figure \ref{figure2} shows the fidelity, $F(t)={\rm Tr}[\rho _{I}(t)\rho (0)]$, as a function of time for Bell initial states. All four Bell initial states result in the same fidelity function up to the fourth decimal place. The solid, dashed, dotted, and dot-dashed lines give the fidelities when the control fields are turned on, using $N=2,3,5,8$, respectively. The insets (a) and (b) show, respectively, the fidelity and concurrence, both calculated at $t=\tau $, as functions of $N$.

\section{conclusion}

In summary, we have shown that it is possible to protect a two-qubit entangled state from disentanglement, using a simple combination of a static field along the $x$ axis and a rotating field in the $yz$ plane. We have tested the method under a very unfavorable circumstance, where dephasing, bit flipping, and dissipation, at a finite temperature, are so effective as to disentangle an unprotected state within a time interval shorter than the characteristic reservoir correlation time, $2\pi /\omega _{c}$, and the thermal correlation time, $\tau_{B}=\beta/\pi$. Even so, the concurrence can be preserved at high fidelity, as shown in Figs. \ref{figure1} and \ref{figure2}. The present result also suggests that it might be possible to protect against disentanglement during the execution of an entangling quantum operation.

\section*{Acknowledgements}

We acknowledge support by Funda\c{c}\~{a}o de Amparo \`{a} Pesquisa do Estado de S\~{a}o
Paulo, Brazil, project number 05/04105-5 and the Millennium Institute for Quantum Information -- Conselho Nacional de Desenvolvimento Cient\'{\i}fico e Tecnol\'{o}gico, Brazil.

\end{document}